\documentclass[reprint, superscriptaddress, amsmath,amssymb, aps, prresearch, longbibliography, floatfix]{revtex4-2}
\usepackage{amsmath}
\usepackage{graphicx}
\usepackage[normalem]{ulem}
\usepackage{bm}
\usepackage{natbib}
\usepackage{dsfont}
\usepackage{tikz}
\usepackage{epsfig}
\usepackage{feynmf}
\usepackage{blindtext, rotating}
\usepackage{mathtools}
\usepackage{dsfont}
\usepackage{subcaption}
\usepackage{physics}
\usepackage{amsfonts}
\usepackage{xcolor}
\usepackage{ragged2e}
\usepackage{siunitx}
\usepackage{comment}
\usepackage{soul}
\usepackage{empheq}
\usepackage{lipsum}
\usepackage{hyperref} 
\hypersetup{breaklinks=true, colorlinks=true, citecolor=blue, linkcolor=cyan, urlcolor=blue,filecolor=blue}
\usepackage{orcidlink}

\usepackage[T1]{fontenc}

\usepackage{xcolor} % Temporario

\DeclareCaptionJustification{justified}{\justifying}

\DeclareSIUnit{\rad}{rad}

\definecolor{bright_blue}{HTML}{85C1E9}
\definecolor{middle_blue}{HTML}{2E86C1}
\definecolor{dark_blue}{HTML}{1B4F72}

\begin{document}

\title{On the quantum nature of strong gravity}
%On the quantum nature of strong gravitational fields
% On the need for quantization in strong field gravity
  % On the quatumness of strong gravitational fields
  % On the quatumness of strong field gravity
  % On the quatum nature of strong field gravity
  % Quantization of weak-field gravity plus the Newtonian limit of Einstein's general relativity requires the quantization of strong-field gravity
  % Weak-field quantum gravity 
  % 

\author{Felipe Sobrero \orcidlink{0009-0009-6497-6612}}
%\email{felipesobrero@cbpf.br}
\affiliation{%
 Centro Brasileiro de Pesquisas Físicas (CBPF), Rua Dr.~Xavier Sigaud 150, Rio de Janeiro 22290-180, RJ, Brazil \\
}

\author{Luca Abrahão \orcidlink{0009-0005-6266-5192}}%
%\email{lucaabrahao99@gmail.com}
\affiliation{Department of Physics, Pontifical Catholic University of Rio de Janeiro, Rio de Janeiro 22451-900, Brazil}

\author{Thiago Guerreiro \orcidlink{0000-0001-5055-8481}}%
\email{thguerreiro@gmail.com}
\affiliation{Department of Physics, Pontifical Catholic University of Rio de Janeiro, Rio de Janeiro 22451-900, Brazil}

\begin{abstract}
 Belenchia et al. \cite{belenchia2018quantum} have analyzed a gedankenexperiment where two observers, Alice and Bob, attempt to communicate via superluminal signals using a superposition of massive particles dressed by Newtonian fields and a test particle as field detector. Quantum fluctuations in the particle motion and in the field prevent signaling or violations of quantum mechanics in this setup. We reformulate this thought experiment by considering gravitational waves emitted by an extended quadrupolar object as a detector for Newtonian tidal fields. %considering that Bob employs an extended quadrupolar object instead of a particle, and uses the gravitational waves emitted by the object's motion as a detector for Alice's tidal fields. 
 We find that quantum fluctuations in the gravitational waves prevent signaling. In the Newtonian limit, rotating black holes behave as extended quadrupolar objects, as consequence of the strong equivalence principle. It follows that consistency of the Newtonian limit of general relativity with quantum mechanics requires the quantization of gravitational radiation, even when the waves originate in strong gravity sources. 
 %In combination with nonlinearities intrinsic to Einstein's gravity, this might lead to interesting non-classical effects in gravitational wave astronomy. 
\end{abstract}

%\keywords{Entanglement, dispersive optomechanics}

\maketitle

\section{Introduction}

Like Maxwell’s electrodynamics, Einstein’s general relativity admits a quantum mechanical description \cite{bronstein2012quantum}. This is the theory of a spin-2 field representing metric perturbations in a background spacetime, and it remains valid in the linearized weak-curvature regime, so long as the spin-2 field carries small energy compared to the Planck scale \cite{donoghue1994general}.
%whose elementary quanta are called gravitons, in analogy to photons, the elementary quanta of light. 

This effective quantum theory of gravity has a number of potentially verifiable predictions. These include gravitational decoherence \cite{blencowe2013effective, de2015decoherence, oniga2016quantum, bassi2017gravitational, lagouvardos2021gravitational}, quantum fluctuations in linearized gravitational waves (GWs) \cite{guerreiro2020gravity, parikh2021b, cho2022quantum, kanno2021noise} and in the gravitational vacuum \cite{ford1995gravitons, ford1996gravitons}, and the possibility of preparing quantum superpositions of massive objects ``dressed'' by Newtonian (Coulombic) fields \cite{feynman2018, marletto2017gravitationally, bose2017spin}. The latter is at the core of the tabletop experimental quantum gravity program \cite{bose2025massive, marletto2025quantum, aspelmeyer2022avoid}, which seeks to observe gravity-mediated entanglement between macroscopic quantum systems, an observation which would rule out classical models of gravity \cite{penrose1996gravity, penrose1996gravity, jacobson1995thermodynamics, verlinde2011origin, oppenheim2023postquantum, carney2025quantum}. 

A central question regarding the quantization of spacetime is whether similar predictions carry over to strong gravitational fields, that is, whether radiative quantum gravitational effects must be present beyond linearized gravity \cite{penrose1976nonlinear}. Our goal in the present work is to show that consistency of quantum mechanics, general relativity and the Newtonian limit of gravity requires a positive answer. 
%By analysing the interplay between superpositions of Newtonian fields, radiative gravitational degrees of freedom, and the Newtonian limit of Einstein's theory, we will show that consistency of quantum mechanics with general relativity requires a positive answer to this question. It is remarkable that such argument for quantizing gravitational waves in the strong field limit can be devised within the linearized regime.

%Our goal in the present work is to show that consistency of quantum mechanics with general relativity requires a positive answer to this question. 

Gravity-mediated entanglement can be described entirely in terms of quantum mechanics in the presence of nonlocal Newtonian gravitational potentials \cite{marletto2025classical}, without explicitly accounting for radiative degrees of freedom. However, by reanalyzing previous gedankenexperiments \cite{mari2016experiments, baym2009two}, Belenchia et al. \cite{belenchia2018quantum} have shown that if quantum superpositions of massive objects dressed by Newtonian fields can be prepared, then a fully consistent quantum description of gravity in the linearized regime requires the
quantization of gravitational radiation and the impossibility of localizing objects to regions smaller than
the Planck length. See also \cite{danielson2022gravitationally, carney2022newton} for similar conclusions from field-theoretic considerations and \cite{danielson2022black, danielson2023killing, gralla2024decoherence, biggs2024comparing} for the related discovery that horizons decohere quantum superpositions.

%this effective quantum theory of weak gravity and its predictions carry over to strong fields? Our goal in the present work is to show that consistency of quantum mechanics with general relativity requires a positive answer.

Here, we investigate whether conclusions similar to those of Belenchia et al. must hold in situations where gravitational fields are strong, in order to maintain consistency between quantum mechanics, Newtonian gravity, and general relativity. We note that a central element of \cite{belenchia2018quantum} are the quantum limitations to measurements of Newtonian gravitational fields using test particles. Instead of test particles, we consider measuring Newtonian tidal fields using GWs emitted by the motion of extended objects with a quadrupole moment. We find that if the GWs are quantized within the effective field theory description of Einstein's gravity, quantum field fluctuations prevent signaling. 
%in order to avoid signaling or violations of quantum complementarity using superpositions of Newtonian fields, the emitted GWs must undergo quantum fluctuations.

In general relativity, rotating black holes (BHs) behave, in the Newtonian limit, as ordinary extended objects with a quadrupole moment, a consequence of the strong equivalence principle \cite{thorne1985laws}. Hence, the consistency of quantum mechanics and general relativity requires that the GWs emitted by the motion of rotating BHs exhibit quantum fluctuations. We conclude that GWs must be quantized, even when they originate from strong-field sources.

%This work is organized as follows. 
In the next Section \ref{belenchia}, we review the Belenchia et al. gedankenexperiment and its apparent paradoxes. Then, we discuss how these paradoxes are resolved by taking into account the quantum mechanical limitations in position measurements of a test particle in Sec. \ref{position-fluctuations}. Next, we introduce a modified version of the gedankenexperiment in Sec. \ref{modified-gedanken}, where we substitute test particles by extended objects with a quadrupole moment, most notably rotating BHs. In Sec. \ref{quadrupoles} we review the dynamics of quadrupolar objects and BHs in Newtonian tidal fields and the corresponding GW coherent states emitted by the motion of those objects. Sec. \ref{paradox-resolution} shows how quantum GW fluctuations are sufficient to resolve the apparent paradoxes discussed previously. 
%As an byproduct of our analysis, we compute the decoherence functional of quadrupole superpositions; this is shown in Sec. \ref{decoherence}. 
We finish in Sec. \ref{Discussion} with a brief discussion on possible consequences of the quantum nature of strong gravity.

We work in natural units where $ \hbar = c = 1 $ keeping factors of $ G = \ell_{P}^{2} $, where $ \ell_{P}$ denotes the Planck length. Roman indices run over spatial coordinates from 1 to 3 and are raised and lowered by the Kronecker delta $ \delta_{ij}$.

%\textcolor{blue}{- Consistency of the far-away (Newtonian + linearized GWs) description of the physics with locality requires the quantization of GWs in the near and strong field zones.}

\section{The Gedankenexperiment
of Belenchia \lowercase{\textit{et al.}}}

\subsection{Test particles as tidal field detectors}\label{belenchia}

%\begin{itemize}
%    \item Can we present this result as a no-go theorem?
%\end{itemize}

In \cite{belenchia2018quantum}, Belenchia et al. consider two observers, Alice and Bob. In a distant past, Alice prepares a massive object with its center of mass in a quantum superposition of two nearby distinct locations $ \vert L \rangle $ and $ \vert R \rangle $,
\begin{eqnarray}
    \vert \Psi \rangle_{A} &=& \frac{1}{\sqrt{2}} \Big[ \vert L\rangle_{A}\vert \gamma_{L}\rangle_{F} +  \vert R\rangle_{A}\vert \gamma_{R}\rangle_{F} \Big] \\ &\approx& \frac{1}{\sqrt{2}} \Big[ \vert L\rangle_{A} +  \vert R\rangle_{A} \Big] \otimes \vert 0 \rangle_{F} \label{superposition1}
\end{eqnarray}
where $ \vert \gamma_{L,R} \rangle_{F}$ denotes the states of the gravitational field produced during Alice's state preparation. We assume Alice prepares her superposition slowly, meaning that in the preparation process, negligible gravitational radiation is emitted. In this case, $ \vert \gamma_{L,R} \rangle_{F}$ are weak coherent states with $ \vert \langle \gamma_{L}\vert \gamma_{R}\rangle \vert \approx 1 $ \cite{breuer2001destruction}, and only a reversible ``false loss of coherence'' occurs in the preparation process \cite{unruh2000false}. 

Far away from Alice, at a distance $ b $, Bob holds a particle in a trap, which he may choose to release or not. The trap is strong enough that if Bob does not turn it off, any influence from Alice's mass is negligible. On the other hand, if Bob chooses to release the particle, it will feel the influence of Alice's mass, and depending on whether her state is $ \vert L \rangle $ or $ \vert R \rangle $, it will evolve towards distinct trajectories after a certain integration time $ T_{B} $. The particle effectively acts as a detector for quantum states, with the state information transmitted by Alice's tidal field.

%After ``integrating'' over a certain time $ T_{B} $, Bob will then be able to distinguish the position of the particle and infer whether Alice's state was $ \vert L \rangle $ or $ \vert R \rangle $.

\begin{figure}[ht!]
    \centering
    \includegraphics[width=\linewidth]{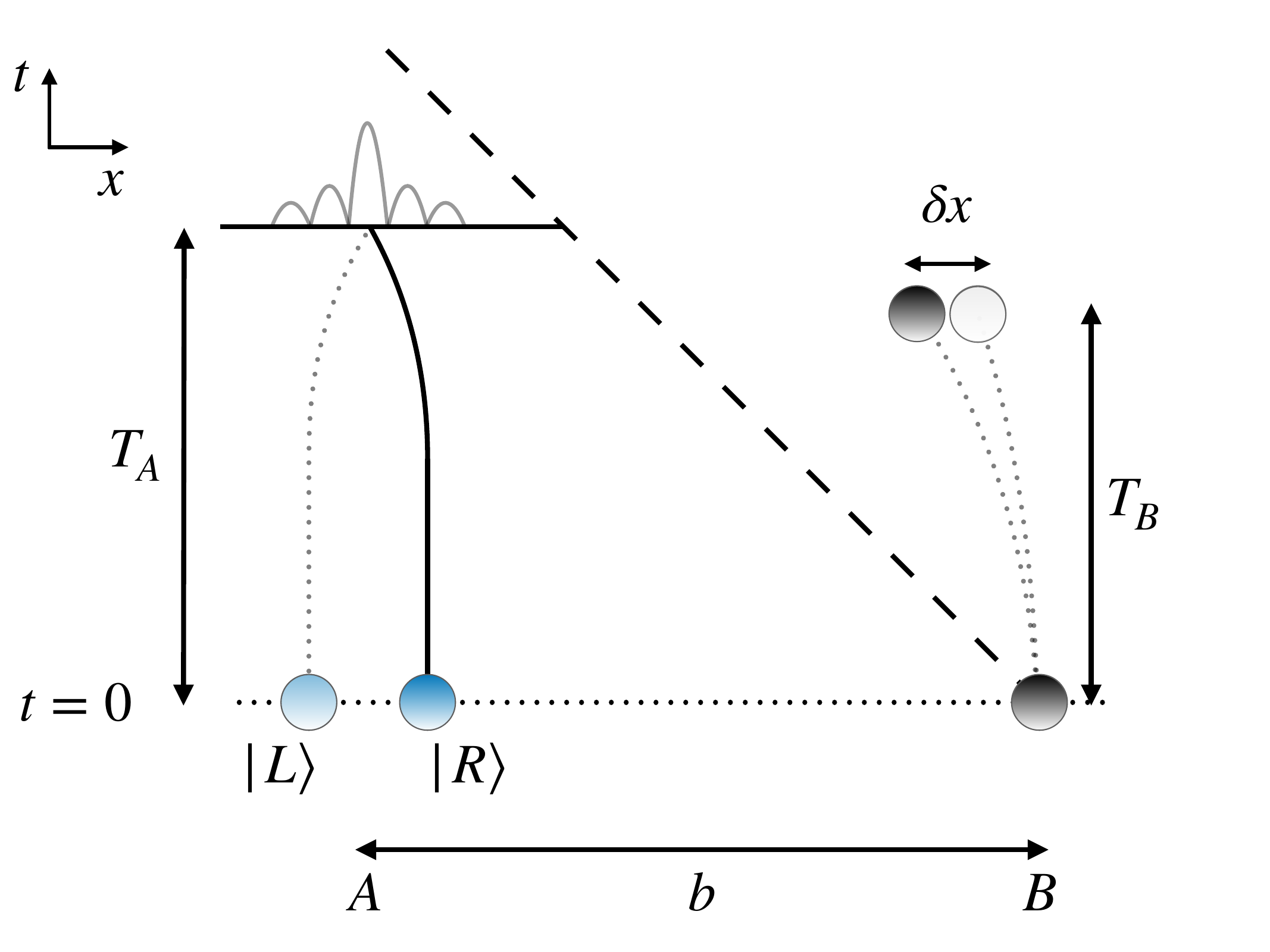}
    \caption{Schematic representation of the Belenchia et al. setup, adapted from \cite{belenchia2018quantum}.
    }
    \label{fig1}
\end{figure}

At a pre-arranged time $ t = 0 $, Bob makes the choice of releasing the particle or not, and Alice begins recombining her wavefunction \eqref{superposition1} to observe interference. We are interested in the case where Alice ``closes her interferometer'' in a time interval 
\begin{eqnarray}
    T_{A} < b \ ,
    \label{Alice_condition_spacelike}
\end{eqnarray}
and Bob observes his test particle over a time interval
\begin{eqnarray}
    T_{B} < b
    \label{Bob_condition_spacelike}
\end{eqnarray}
Under these conditions, Alice and Bob remain spacelike separated throughout the entire experiment. A schematic representation of the gedankenexperiment is shown in Fig. \ref{fig1}.

%\textit{The classical picture.} 
Now comes the paradox. If Bob chooses to release the particle, he will in principle be able to obtain which-path information on Alice's state at spacelike separation. By the principle of complementarity, this implies Alice's superposition must undergo decoherence. This implies a violation of causality, since the purity of Alice's state now depends on Bob's choice of releasing the particle or not, eventough both observers are spacelike separated. Alternatively, if Bob obtains information on Alice's state without causing any loss of coherence, causality is maintained, but the principle of complementarity is violated. It seems either causality, quantum mechanics, or both, are violated to some degree.

\subsection{Position quantum fluctuations to the rescue}\label{position-fluctuations}
%\textit{Quantum mechanics to the rescue.} 

As pointed out in \cite{belenchia2018quantum}, the paradox is resolved by noting that quantum fluctuations of the spacetime metric impose fundamental limits on the measurement of the position of a particle. 
%To resolve the paradox, note that spacetime itself undergoes quantum fluctuations which limits Bob's ability of measuring the position of his test particle.

The components of the Riemann curvature tensor averaged over a region of size $ \ell $ undergo vacuum fluctuations on the order of $ \vert \Delta R_{\alpha\beta\gamma\delta}\vert \sim \ell_{P}/\ell^{3}$ \cite{wheeler1957nature}. Integrating the geodesic deviation equation, we find that the relative position of two objects has a minimum fluctuation of $ \Delta x \sim \ell_{P} $; such fluctuations can be rigorously derived in the effective quantization of Einstein gravity \cite{guerreiro2020gravity, parikh2021b, cho2022quantum, kanno2021noise}. Over a time $ T_{B} $, Bob's particle undergoes a displacement $ \delta x \sim GQ_{A} T_{B}^{2} / b^{4}$, where $ Q_{A} $ denotes the effective mass quadrupole moment of Alice's superposition. The condition that Bob obtains which-path information on Alice's state then becomes $ \delta x > \Delta x$, or
\begin{eqnarray}
    \frac{GQ_{A}}{b^{4}}T_{B}^{2} > \ell_{P} \ ,
\end{eqnarray}
which can be written as 
\begin{eqnarray}
    \left(\frac{\ell_{P}}{b}\right) \left(\frac{Q_{A}}{b}\right) \left(\frac{T_{B}}{b}\right)^{2} > 1 \ . 
    \label{bob_gains_which_path_belenchia}
\end{eqnarray}

We now turn to the condition that Alice's state does not undergo decoherence during the experiment. While closing her interferometer, Alice emits GWs in coherent states with a mean number of gravitons $N$ and a characteristic frequency $ \omega_{A} \sim 1/T_{A} $, where $ T_{A}$ is the time over which she recombines the components of her wavefunction. The GWs emitted carry away information on her state, resulting in loss of coherence \cite{breuer2001destruction}. To avoid decoherence, the mean number of gravitons emitted during the experiment must satisfy $ N < 1$. Since coherent states are non-orthogonal, this guarantees that negligible information on Alice's state leaks to the gravitational degrees of freedom. 

The mean number of gravitons emitted by Alice can be computed from Einstein's quadrupole formula \cite{ligo2017basic}, 
\begin{eqnarray}
    P = \frac{dE}{dt} \approx \frac{G}{5}\langle \dddot{Q}^{A}_{ij}\dddot{Q}^{A}_{ij} \rangle \sim G Q_{A}^{2} \omega_{A}^{6} \ .
    \label{quadrupole_formula}
\end{eqnarray}
We have $ N \sim E / \omega_{A} \sim PT_{A}/\omega_{A}$, yielding
\begin{eqnarray}
    N \sim G \left(  \frac{Q_{A}}{T_{A}^{2}}\right)^{2} \ .
\end{eqnarray}
The condition that Alice does not decohere becomes,
\begin{eqnarray}
    Q_{A} < \frac{T_{A}^{2}}{\ell_{P}} \ 
    \label{Alice_no_decoherence}.
\end{eqnarray}

Combining \eqref{bob_gains_which_path_belenchia} and \eqref{Alice_no_decoherence}, Bob will obtain which-path information and Alice will avoid decoherence if
\begin{eqnarray}
    \left(\frac{T_{A}}{b}\right)^{2} \left(\frac{T_{B}}{b}\right)^{2} > 1 \ . 
    \label{Belenchia_final_condition}
\end{eqnarray}
This is incompatible with the spacelike separation conditions \eqref{Alice_condition_spacelike} and \eqref{Bob_condition_spacelike}, which resolves the paradox: no which-path information from Alice's state can be obtained by Bob at spacelike separation. Therefore, no violations of causality or complementarity can occur.  

%Moreover, if Bob does obtain information on the state, when Alice and Bob are timelike separated, Alice's state will undergo decoherence in accordance with complementarity, it is ... but this decoherence 

%We refer the reader to \cite{belenchia2018quantum} for a detailed discussion on the gedankenexperiment. 

\subsection{GWs from the motion of black holes as tidal field detectors}\label{modified-gedanken}

We now want to introduce a modified version of the Belenchia et al. gedankenexperiment, where Bob uses gravitational waves emitted by a strong-field source, rather than particles, as a detector for Alice's superposition state.

\begin{figure}[ht!]
    \centering
    \includegraphics[width=\linewidth]{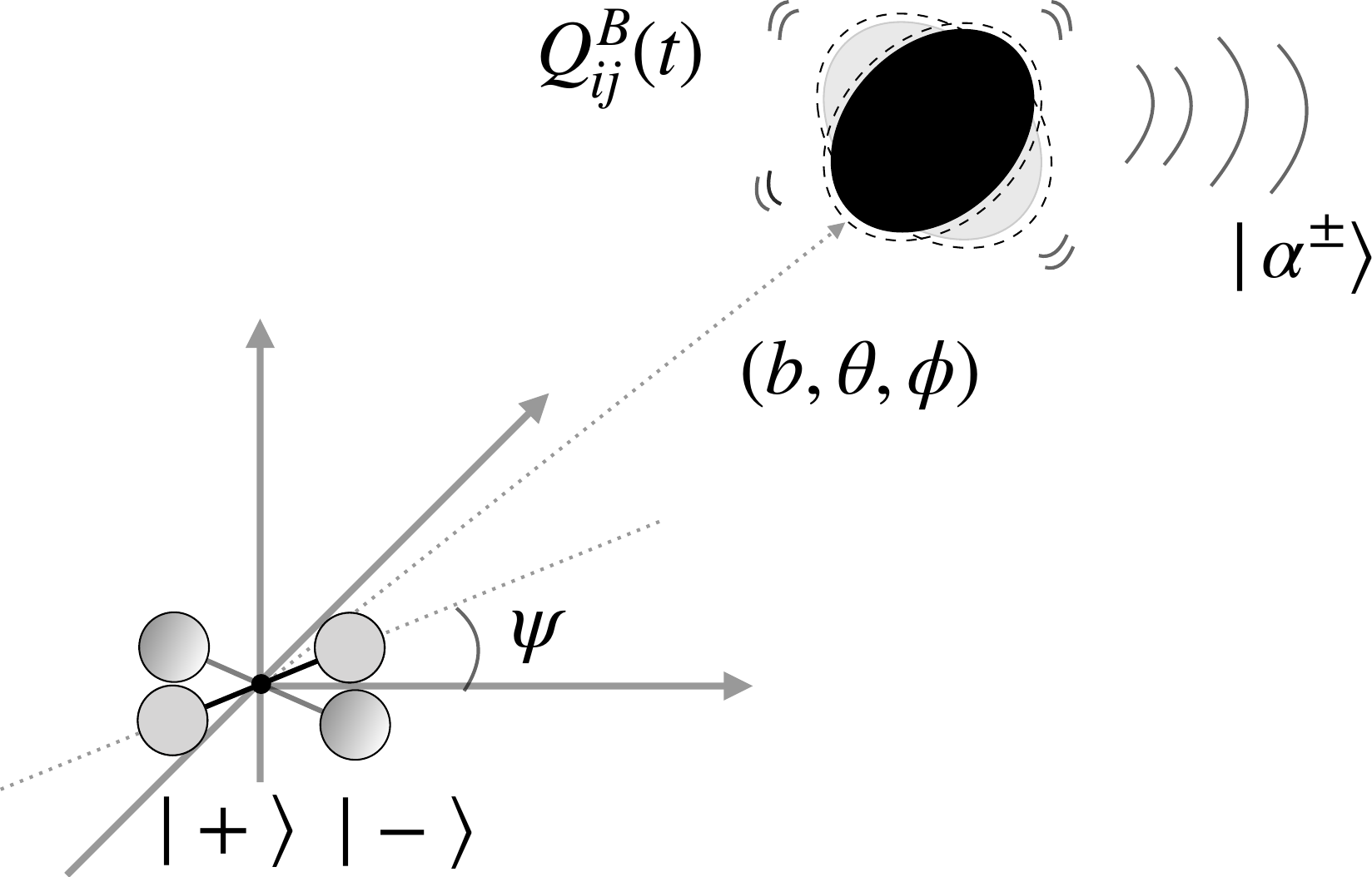}
    \caption{Modified version of the gedankenexperiment, where we consider tidal fields produced by Alice's binary mass system placed in the $xy$ plane. The binary mass system is prepared in a superposition of distinct orientations $\vert \pm \rangle$,  corresponding to angles $\pm \psi$ with respect to the $x$ axis. To detect the superposition, Bob measures gravitational waves emitted by the motion of a rotating black hole at position $\mathbf{x} = (b,\theta,\phi)$, with quadrupole moment $Q_{ij}^{B}(t)$. Depending on Alice's tidal field, the waves are emitted in coherent states $\vert \alpha^{\pm} \rangle$; Bob can obtain which-path information on Alice's state when $\vert\langle \alpha^{+}\vert\alpha^{-}\rangle\vert \approx 0$.
    }
    \label{fig2}
\end{figure}

From now on, Alice's setup consists of a binary mass system placed in a superposition of distinct \textit{orientations}, 
\begin{eqnarray}
    \vert \Psi' \rangle_{A} &=& \frac{1}{\sqrt{2}} \Big[ \vert +\rangle_{A}\vert \gamma_{+}\rangle_{F} +  \vert -\rangle_{A}\vert \gamma_{-}\rangle_{F} \Big] \\
    &\approx& \frac{1}{\sqrt{2}} \Big[ \vert +\rangle_{A} +  \vert -\rangle_{A} \Big] \otimes \vert 0 \rangle_{F} \label{superposition2}
\end{eqnarray}
where $ \vert \pm \rangle_{A} $ denotes the orientation states of the binary in the $ xy $ plane given by angles $ \pm \psi $ with respect to the $ x $ axis, and $\vert \gamma_{\pm} \rangle_{F}$ denotes the states of the gravitational field resulting from Alice's preparation procedure. As before, we assume Alice prepares her superposition slowly, such that only weak coherent states are emitted, with $ \vert \langle \gamma_{+} \vert \gamma_{-} \rangle \vert \approx  1$ \cite{breuer2001destruction} and only a reversible ``false loss of coherence'' can occur at intermediate times \cite{unruh2000false}. Alice's  system has total mass $ m $ and separation between the masses $ d $.

Alice's superposition effectively produces a tidal field, which we will discuss in Sec. \ref{sec: precession} below. As before, Bob wants to measure this tidal field, but instead of employing a test particle to do that, he uses an extended quadrupolar object. We will treat the motion of Bob's quadrupole within the Newtonian limit, which is a good approximation when it moves slowly and 
is far from other sources of gravity. Again, we consider Bob's quadrupole detector rests at a distance $ b $ from Alice, at a coordinate location parametrized by a polar and azimuthal angles $\theta,\phi$.

In the Newtonian limit, far from other sources of gravity, rotating (Kerr) BHs behave as ordinary extended objects with a quadrupole moment. We will from now on consider Bob's quadrupole to be a rotating BH. 
%Examples of extended objects are Kerr black holes (BHs), and henceforth we will treat Bob's object as such.
We assume Bob has control over the BH and can choose to ``trap'' it, preventing it from moving appreciably in the presence of the external field, or releasing it to move freely if he so chooses \footnote{Trapping the BH is not  an essential requirement, Bob could instead influence its initial condition, such as the direction of the spin vector at $t=0$.}. 

In order to detect the motion of the BH, Bob rests well within its wave zone \cite{arnowitt1961wave}, and detects changes in the object's orientation by measuring the GWs emitted by its motion. In this far away description, the BH follows the strong equivalence principle and can be modeled as a quadrupole moment in a Newtonian tidal field \cite{thorne1985laws}. We discuss the laws of motion for this system shortly below. %Similarly, the emitted GWs are treated linear perturbations. 

As before, we are interested in the case that the duration of Alice's experiment $T_{A}$, and the time $ T_{B} $ over which Bob observes the GWs emitted by his ``BH tidal detector'' satisfies the spacelike separation conditions \eqref{Alice_condition_spacelike} and \eqref{Bob_condition_spacelike}. A Schematic depiction of the setup can be seen in Fig.~\ref{fig2}. 

Once again, now comes the paradox. The motion of Bob's BH will depend on Alice's states $ \vert \pm \rangle_{A} $, through their corresponding tidal fields. Consequently, GWs emitted by the motion of the BH will be distinct for each orientation. If GWs were classical, it would in principle be possible to obtain which-path information on Alice's superposition state at spacelike separation, leading to violations of causality, complementarity, or both. 

We can expect the same conclusions would hold independently of whether Bob's detector is composed of black holes and gravitational waves (which are pure spacetime degrees of freedom), or any other from of mass-energy, which places theories were some systems behave quantum mechanically while others are inherently classical in a delicate position.

\section{GWs from the motion of quadrupoles in Newtonian fields}\label{quadrupoles}

Our aim in the coming sections is to show that the paradox can be avoided by considering quantum fluctuations in the gravitational radiation emitted by the motion of Bob's BH. To proceed with that, we must first describe the dynamics of the system in more detail.

\subsection{Action}

Consider the dynamics of an object with quadrupole moment $ Q_{ij}^{B} $ (which we take to be Bob's BH) in the presence of a Newtonian tidal field $ \mathcal{E}_{ij}^{A} $, produced by Alice's binary mass system. We also consider the coupling of GW perturbations $ h_{ij} $ to Bob's quadrupolar object. Far from the objects' surface (the BH horizon) and in the Newtonian limit, the system is described by the total effective action
\begin{align}
    S = \frac{1}{2} \int dt \left[ \left( \mathcal{E}_{ij}^{A} + \ddot{h}_{ij} \right) Q_{ij}^{B} + \frac{1}{4\kappa^{2}} \int d^{3}\mathbf{x} \ \partial_{\mu}h_{ij}\partial^{\mu}h^{ij} \right] \ , 
    \label{action}
\end{align}
where $ \kappa \equiv \sqrt{8\pi G}$, the first two terms represent the quadrupole interaction with the external tidal and GW fields, and the third term is the free action for the GW perturbations in the TT gauge, where $ \partial_jh_{ij} = h_{ii} = 0 $ and $\mu $ runs from 0 to 3. 

We work in a cubic box of volume $ V = L^{3} $, and decompose GW perturbations in Fourier modes
\begin{equation}
    h_{ij}(t,\mathbf{x}) = \frac{2\kappa}{\sqrt{V}} \sum_{\mathbf{k},\lambda} h^{\lambda}_\mathbf{k}(t) e^{i \mathbf{k} \cdot \mathbf{x}} e^{\lambda}_{ij}(\mathbf{k}) \ ,
    \label{fourier_modes}
\end{equation}
where $ e^{\lambda}_{ij}(\mathbf{k}) $ is the polarization tensor satisfying $ e^{\lambda}_{ij}(\mathbf{k}) e^{ij}_{\lambda'}(\mathbf{k}) = \delta_{\lambda \lambda'} $ (with $\lambda = +, \times$), $ k_{i} = 2\pi n_{i}/L $ ($ n_{i} \in \mathbb{Z} $)  are the components of the wave-vector $\mathbf{k}$, and reality of the fields implies $ h^{\lambda}_{-\mathbf{k}}e^{\lambda}_{ij}(-\mathbf{k}) = h^{\lambda *}_{\mathbf{k}}e^{\lambda}_{ij}(\mathbf{k}) $. Substituting \eqref{fourier_modes} in \eqref{action}, using the Fourier orthogonality relation and performing integration by parts neglecting boundary terms, we rewrite the action as $ S = S_{0} + S_{N} + S_{I} $, where
\begin{eqnarray}
    S_{0} = \frac{1}{2}\int dt \ \sum_{\lambda,\mathbf{k}}\left(\dot{h}^{\lambda}_{\mathbf{k}} \dot{h}^{\lambda*}_{\mathbf{k}}-\mathbf{k}^2 h^{\lambda}_{\mathbf{k}}h^{\lambda*}_{\mathbf{k}} \right)
\end{eqnarray}
is the free GW perturbation term, 
\begin{eqnarray}
    S_{N} = \int dt \ \mathcal{E}_{ij}^{A}Q_{ij}^{B} \equiv \int dt \ \mathrm{Tr}\left( \mathcal{E}^{A} Q^{B} \right)
    \label{quadrupole_energy}
\end{eqnarray}
is the interaction with the Newtonian tidal field, and
\begin{eqnarray}
    S_{I} = \int dt \ \frac{2\kappa}{\sqrt{V}}\sum_{\lambda}\sum_{\mathbf{k}\leq\Omega_m} h^{\lambda}_{\mathbf{k}}(t)e^{\lambda}_{ij}(\mathbf{k}) \ddot{Q}^{B}_{ij}
    \label{GW_interaction}
\end{eqnarray}
describes the interaction of Bob's quadrupole with GW perturbations. Note we have introduced the cut-off frequency $ \Omega_{m} $, restricting momentum sums and imposing that the quadrupole cannot emit GWs of arbitrarily high frequency. We take the cut-off frequency to be given by the light crossing limit $\Omega_{m} R \lesssim 1$, where $  R \sim 2 GM $ is the dimension of the BH, yielding
\begin{eqnarray}
    \Omega_{m} \lesssim (2GM)^{-1} \ .
\end{eqnarray}

We quantize the GW perturbations by writing the Fourier components of the field $ h^{\lambda}_{\mathbf{k}}(t)$ in the interaction picture $ h^{\lambda}_{I,\mathbf{k}}(t) $ in terms of annihilation and creation operators \cite{cohen1997photons},
\begin{equation}
    h^{\lambda}_{I,\mathbf{k}}(t) = a^{\lambda}_\mathbf{k} u_{\mathbf{k}}(t) + a^{\lambda \dagger}_{-\mathbf{k}} u^{*}_{\mathbf{k}}(t) \label{free_field_amplitude}
\end{equation}
where $ u_{\mathbf{k}}(t) $ denotes mode functions and $ a^{\lambda}_\mathbf{k}, a^{\lambda \dagger}_{\mathbf{k}} $ satisfy the bosonic commutation relations
\begin{equation}
[a^{\lambda}_\mathbf{k},a^{\lambda'\dagger}_\mathbf{k'}] = \delta_{\mathbf{k},\mathbf{k}'}\delta_{\lambda,\lambda'} \ , ~[a^{\lambda}_\mathbf{k},a^{\lambda'}_\mathbf{k'}]=[a^{\lambda\dagger}_\mathbf{k},a^{\lambda'\dagger}_\mathbf{k'}]=0 \ , \label{eq: commutation relation}
\end{equation} and $a^{\lambda}_{\mathbf{k}}\ket{0}=0 ~\forall \mathbf{k}, \lambda$ defines the vacuum state $\ket{0}$. 
%Mode functions are normalized according to the scalar product \cite{fulling1989aspects}.
\begin{comment}
    \begin{equation}
    \langle \phi,\psi \rangle =-i \int \sqrt{-g} ~n^\mu d\Sigma \Big(\phi \partial_\mu \psi^*-\psi^*\partial_\mu \phi\Big) \ , 
\end{equation}
where $ n^\mu $ is a future-directed unit vector orthogonal to a spacelike hypersurface $ \Sigma$ with volume element $d\Sigma$. 
%The scalar product is independent of $\Sigma$ \cite{fulling1989aspects}.
\end{comment} 
For the Minkowski vacuum mode functions are given by
\begin{eqnarray}
    u_\mathbf{k} = e^{-ikt} / \sqrt{2k} \ , 
\end{eqnarray}
where $ k \equiv \vert \mathbf{k}\vert$. 
\begin{comment}
and
\begin{equation}
    \langle u_\mathbf{k},u_{\mathbf{k}'} \rangle=\delta_{\mathbf{k},\mathbf{k}'} \ .
\end{equation}
\end{comment}

\subsection{GW emission}\label{sec: gw emission}

%We can compute the emission of GWs by writing the Eqs. of motion for operators $ h_{\mathbf{k}}^{\lambda}$. 

Varying the action with respect to $ h_{\mathbf{k}}^{\lambda *} $ and using the field reality conditions we find that GWs are sourced by second derivatives of the quadrupole moment,
\begin{equation}
\ddot{h}^{\lambda}_{\mathbf{k}}+\mathbf{k}^2h^{\lambda}_{\mathbf{k}}=\frac{\kappa}{\sqrt{V}}e^{\lambda*}_{ij}(\mathbf{k})\ddot{Q}^{B}_{ij} \label{GW_eq} \ ,
\end{equation}
which is formally solved by~\footnote{We can formally assume the GW and quadrupole interaction is switched on adiabatically $ \kappa \rightarrow \kappa \epsilon(t)$ at $t = 0$, where $ \epsilon(t) $ is a switching function with $ \epsilon(t \leq -\delta) = \epsilon(t \geq T + \delta) = 0 $,  $ \epsilon(0 \leq t \leq T) = 1 $ and $ \delta $ represents a switching time-scale. Assuming $\delta $ small with respect to any other time-scale in the problem, we neglect corrections arising from this switching time-scale.}, 
\begin{equation}
    h^{\lambda}_{\mathbf{k}}(t) = h^{\lambda}_{I,\mathbf{k}}(t) + \frac{\kappa}{\sqrt{V}}e^{\lambda*}_{ij}(\mathbf{k})\int_{0}^{t}dt' \frac{\sin k(t-t')}{k}\ddot{Q}^{B}_{ij}(t') \ .\label{GW_formal_solution}
\end{equation}

\noindent We see the effect of the quadrupole $ Q_{ij}^{B}(t) $ is to add a c-number to the operator $ h_{\mathbf{k}} $, indicating that GWs are emitted in a coherent state as a consequence of the linear coupling in \eqref{GW_interaction} \cite{unruh2000false}. Defining displacement operators $ D_{\mathbf{k}}^{\lambda}(\alpha_{\mathbf{k}}^{\lambda})$ for modes $ (\mathbf{k},\lambda) $, we can write \eqref{GW_formal_solution} as
\begin{eqnarray}\label{displacement}
    h^{\lambda}_{\mathbf{k}}(t) 
 &=& D^{\lambda}_{\mathbf{k}}(\alpha^{\lambda}_\mathbf{k}) h^{\lambda}_{I,\mathbf{k}}(t) D^{\lambda \dagger}_{\mathbf{k}}(\alpha^{\lambda}_\mathbf{k}) \nonumber \\ &=& \left( a^{\lambda}_\mathbf{k} + \alpha^{\lambda}_\mathbf{k} \right) u_{\mathbf{k}}(t) + a^{\lambda \dagger}_{-\mathbf{k}} u^{*}_{\mathbf{k}}(t) \ , 
\end{eqnarray}
where 
\begin{equation}
    \alpha^{\lambda}_\mathbf{k}(t) = \frac{\kappa}{\sqrt{V}}e^{-i k t} e^{\lambda*}_{ij}(\mathbf{k}) \int_{0}^{t} dt' \sqrt{\frac{2}{k}} \sin k(t-t') \ddot{Q}^{B}_{ij}(t') \label{alpha}
\end{equation}
is the amplitude for a coherent state in mode $(\mathbf{k},\lambda)$, 
\begin{eqnarray}
    \vert \alpha^{\lambda}_\mathbf{k}(t)\rangle = D^{\lambda}_{\mathbf{k}}(\alpha^{\lambda}_\mathbf{k}) \vert 0 \rangle \ .
\end{eqnarray}
Taking into account all modes, the quadrupole moment emits GWs in the coherent state,
\begin{eqnarray}
    \vert \alpha(t) \rangle = \bigotimes_{\mathbf{k},\lambda} \vert \alpha^{\lambda}_\mathbf{k}(t)\rangle \ .
\end{eqnarray}

%Given two distinct quadrupole source configurations $ Q_{ij}^{B+}(t)$ and $ Q_{ij}^{B-}(t)$ produced by the interaction of Bob's BH with Alice's tidal field

We now consider two distinct quadrupole configurations $ Q_{ij}^{B+}(t)$ and $ Q_{ij}^{B-}(t)$, corresponding to each of the tidal fields produced by Alice's $\vert \pm\rangle_{A}$ states. The overlap between the coherent GW states emitted by each of the two quadrupole configurations reads
\begin{eqnarray}
    \left|\langle{\alpha^{+}(t)}|{\alpha^{-}(t)}\rangle\right|^2 &=& \prod_{\mathbf{k},\lambda} \left|\langle{\alpha^{\lambda+}_\mathbf{k}(t)}|{\alpha^{\lambda-}_\mathbf{k}(t)}\rangle\right|^2 \nonumber \\ &=& e^{-D(t)} \ ,  \label{eq: alpha + and - expected value}
\end{eqnarray}
where each GW mode contributes as 
\begin{equation}
\left|\langle{\alpha^{\lambda+}_\mathbf{k}(t)}|{\alpha^{\lambda-}_\mathbf{k}(t)}\rangle\right|^2 = e^{-\left| \alpha^{\lambda+}_\mathbf{k}(t)- \alpha^{\lambda-}_\mathbf{k}(t)\right|^2} \ ,
\end{equation}
and we have defined the \textit{decoherence function} \cite{dowker1992quantum},
\begin{eqnarray}
    D(t) = \sum_{\mathbf{k},\lambda} \vert  \alpha^{\lambda+}_\mathbf{k}(t)- \alpha^{\lambda-}_\mathbf{k}(t)\vert^{2} \ , \label{decoherence_functional}
\end{eqnarray}
where $\alpha_{\mathbf{k}}^{\lambda \pm}$ are given by \eqref{alpha}. Recall that sums over wave-vectors are limited by $ \vert \mathbf{k} \vert \leq \Omega_{m} $. 
%Our goal is to compute $ D(t) $.

Invoking the polarization identity,
\begin{equation}
    \sum_\lambda e^{\lambda*}_{ij}(\mathbf{k})e^{\lambda}_{k \ell}(\mathbf{k}) = \frac{1}{2}\left( P_{ik} P_{j\ell} + P_{i\ell} P_{jk} - P_{ij} P_{k\ell} \right),
\end{equation}
where $P_{ij} = \delta_{ij} - k_i k_j/k^2$, promoting sums to the continuum limit $ \sum_{\mathbf{k}} \rightarrow \frac{V}{(2\pi)^{3}}\int d^{3}\mathbf{k}$, with $d^3 \mathbf{k} = k^2 dk ~d\Omega$, and using the angular integral 
\begin{eqnarray}
    \int d\Omega \left( P_{ik} P_{j\ell} + P_{i\ell} P_{jk} - P_{ij} P_{k\ell} \right) \nonumber \\ = \frac{8\pi}{5}\left( \delta_{ik} \delta_{j\ell}+ \delta_{i\ell} \delta_{jk} -\frac{2}{3} \delta_{ij} \delta_{k\ell} \right) \ ,
\end{eqnarray}
we can write the decoherence function as
\begin{eqnarray}
     D(t)  &=& \frac{4\kappa^2}{5 \pi^2} \int_{0}^{\Omega_m}dk k \  \nonumber  \\ 
     && \times\int_{0}^{t}\int_{0}^{t} dt' dt'' \sin k(t-t') \sin k(t-t'') g(t',t'') \nonumber\\
     \label{coherent_state_integration}
\end{eqnarray}
where
\begin{equation}
    g(t',t'') = \frac{d^2}{dt'^2}\Delta\hat{Q}^{B}_{ij}(t')\frac{d^2}{dt''^2}\Delta\hat{Q}^{B ij}(t'')
    \label{gtautau}
\end{equation}
and
\begin{equation}
    \Delta{Q}^{B}_{ij}(t) \equiv {Q}^{B+}_{ij}(t) - {Q}^{B-}_{ij}(t).
    \label{delaQ}
\end{equation}

\subsection{Precession of black holes}\label{sec: precession}

The Newtonian interaction between Bob's quadrupole moment $Q^{B}$ and Alice's tidal field is given by the action term \eqref{quadrupole_energy}, which is extremized when $ \mathcal{E}^{A} $ and $ Q^{B} $ share common eigenaxes, that is, when the quadrupole and tidal field ``align''. This follows from von Neumann's trace inequality \cite{horn1994topics}. At the level of equations of motion \cite{thorne1986black}, the quadrupole feels a torque $ N_{i} $ given by \footnote{We neglect the effect of $\ddot{h}_{ij}$ in the quadrupole equation of motion, as this term turn only contributes to higher orders in $D(t)$.},
\begin{eqnarray}
    N_{i} = -\epsilon_{ijk} \mathcal{E}^{A}_{kl} Q^{B}_{jl} \ .
    \label{quadrupole_motion}
\end{eqnarray}

Explicitly, the Lagrangian associated to \eqref{quadrupole_energy} reads \cite{poisson2014gravity},
\begin{eqnarray}
    \mathrm{Tr}(\mathcal{E}^{A}Q^{B}) = -\frac{3G}{2b^{5}} &&\left[    2\mathrm{Tr}(Q^{A}Q^{B}) - 20 \left( \hat{x}^{T} \cdot Q^{A}Q^{B} \cdot \hat{x}\right) \right. \nonumber \\ 
    &&\left. ~~~~~+ 35 \left( \hat{x}^{T}\cdot Q^{A} \cdot \hat{x} \right) \left( \hat{x}^{T}\cdot Q^{B} \cdot \hat{x} \right)  \right] \nonumber \\
    \label{newtonian-lagrangian}
\end{eqnarray}
where $ \hat{x} = \hat{\mathbf{x}}/b $ denotes Bob's normalized position vector. In the situation depicted in Figure \ref{fig2}, Alice's quadrupole has the form
\begin{equation}
    Q^{A} = \Bar{Q}^{A} \hat{Q}^{A} = \Bar{Q}^{A} \begin{pmatrix}
\hat{Q}_{xx} & \hat{Q}_{xy} & 0 \\
\hat{Q}_{xy} & \hat{Q}_{yy} & 0 \\
0 & 0 & \hat{Q}_{zz}
\end{pmatrix}
\end{equation}
with 
\begin{equation}
    \Bar{Q}^{A} = md^{2}
\end{equation}
and
\begin{eqnarray}
    \hat{Q}_{xx} &=& 1 + 3\cos 2\psi \label{eq: Qxx}\\  
    \hat{Q}_{yy} &=& 1 - 3\cos 2\psi \label{eq: Qyy}\\  
    \hat{Q}_{xy} &=& 3 \sin 2\psi \label{eq: Qxy}\\  
    \hat{Q}_{zz} &=& -2 \label{eq: Qzz}
\end{eqnarray}
where we recall $\psi$ denotes the orientation of the quadrupole in the $xy$ plane. We define the dimensionless tidal field
\begin{eqnarray}
    \hat{\mathcal{E}}^{A}_{ij} = \frac{3}{2} \left[ \hat{Q}_{ij}^{A} - 5\left(   \hat{Q}^{A}_{ik} \hat{x}^{k}\hat{x}_{j} + \hat{Q}^{A}_{jk} \hat{x}^{k}\hat{x}_{i} \right) \right. \nonumber \\ \left. 
 + \frac{35}{2} \left( \hat{Q}^{A}_{kl} \hat{x}^{k}\hat{x}^{l}  \right) \hat{x}_{i}\hat{x}_{j}  \right]
    \label{dimensionless-tidal-field}
\end{eqnarray} 
and the coupling constant
\begin{eqnarray}
    g = \frac{2G\Bar{Q}_A}{b^{5}}
\end{eqnarray}
such that the tidal field in the interaction Lagrangian \eqref{newtonian-lagrangian} reads
\begin{eqnarray}
    \mathcal{E}^{A}_{ij} =- g \hat{\mathcal{E}}^{A}_{ij} \ .
\end{eqnarray}
Note that the tidal field depends on Alice's quadrupole orientation $\psi$, as well as on the polar and azimuthal angles of Bob's position vector $(\theta,\phi)$, i.e. $ \hat{\mathcal{E}}^{A} = \hat{\mathcal{E}}^{A}(\psi,\theta,\phi) $. We refer to Appendix \ref{sec: appendix} for an explicit expression of the tidal field at an arbitrary position in space.

For a rotating BH of mass $ M$ and spin $ J_{i} = Ma \hat{J}_{i} $, with $ \hat{J}_{i} $ the unit vector pointing along the spin direction and $ a $ the length scale associated to the hole's angular momentum (which is related to the dimensionless angular momentum number $\chi = a/GM$, with $ 0 \leq \chi \leq 1$), the quadrupole moment reads, 
\begin{eqnarray}
    Q^{B}_{ij} = -M a^2 \hat{Q}^{B}_{ij}, ~\hat{Q}_{ij}^{B} = \hat{J}_i \hat{J}_j - \frac{1}{3}\delta_{ij} \ .
    \label{BH_quadrupole}
\end{eqnarray}
Introducing the characteristic frequency 
\begin{eqnarray}
    \omega = \chi GM g = 2\chi \frac{G^{2}M\bar{Q}_{A}}{b^{5}}
\end{eqnarray}
and the rescaled time $\tau = \omega t$, we may rewrite \eqref{quadrupole_motion} as,
\begin{eqnarray}
    \dot{\hat{J}} = - (\hat{\mathcal{E}}^{A}\hat{J}) \times \hat{J}
    \label{eq-motion-final}
\end{eqnarray}
where from now on, dots denote ``$d/d\tau$''. This equation describes a torqued precession of the spin of BHs in the presence of distant Newtonian sources of gravity \cite{thorne1985laws}. Note that the magnitude of the BH's angular momentum is conserved under Eq. \eqref{eq-motion-final},
\begin{eqnarray}
    \frac{d}{d\tau}\vert \hat{J}\vert^{2} = 2 \hat{J} \cdot \dot{\hat{J}} = - 2 \hat{J} \cdot \left(  (\hat{\mathcal{E}}^{A}\hat{J}) \times \hat{J} \right) = 0 
\end{eqnarray}
and hence $ \vert \hat{J}(t)\vert = 1 \ \forall \tau$; importantly for us, the magnitude of the BH's angular momentum remains bounded for all times.

\begin{figure}[ht!]
    \centering
    \includegraphics[width=\linewidth]{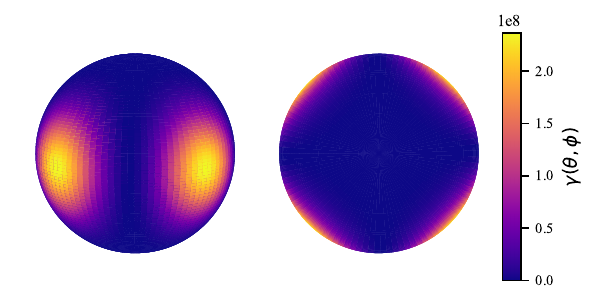}
    \caption{Plot of $\gamma = \gamma(\theta,\phi)$ as a function of Bob's polar and azimuthal angles $ \theta,\phi $, for fixed orientation of Alice's superposition state $\psi = \pi/4$; left: lateral view, right: top view.  
    }
    \label{fig-final}
\end{figure}

\section{GW quantum fluctuations
to the rescue}\label{paradox-resolution}

We now have all the elements to reanalyze the gedankenexperiment described in Sec. \ref{modified-gedanken}, taking into account quantum  fluctuations present in the GWs emitted by the motion of Bob's BH under the influence of Alice's tidal field.

Once again, there are two conditions that must be met in order to have paradoxes. First, in the course of Alice's interference experiment the binary system must be moved slowly, so that her superposition undergoes negligible decoherence. In our case, this yields the same condition as in Eq. \eqref{Alice_no_decoherence}. Second, Bob must be able to distinguish the motion of his BH under each of the tidal field configurations produced by Alice's $ \vert \pm \rangle_{A}$ states. Since Bob is measuring the GWs emitted by the motion of the BH, the condition for distinguishing $ \vert + \rangle_{A} $ and $ \vert - \rangle_{A} $ is $ \vert \langle \alpha^{+}(T_{B}) \vert \alpha^{-}(T_{B})\rangle \vert < 1$, or,
\begin{align}
    D(T_{B}) > 1 \ .\label{Bob-no-decoherence}
\end{align}
We will now show that \eqref{Alice_condition_spacelike}, \eqref{Bob_condition_spacelike}, \eqref{Alice_no_decoherence} and \eqref{Bob-no-decoherence} cannot be met simultaneously.

%\subsection{Decoherence function}

We introduce the dimensionless cutoff 
\begin{eqnarray}
    \Lambda = \Omega_{m} / \omega = \frac{1}{2GM\omega}
\end{eqnarray} 
In the limit of large $\Lambda$, the decoherence function can be written as
\begin{equation}
    D(\tau) = D_{0} F(\Lambda,\tau)
    \label{decoherence-dimensionless}
\end{equation}
with 
\begin{equation}
    D_{0} = \frac{\chi^{4}}{20\pi}\left(\frac{R_{s}}{\ell_{P}}\right)^{2}\frac{1}{\Lambda^{4}}
    \label{D0}
\end{equation}
and 
\begin{eqnarray}
    F(\Lambda,\tau) \approx \frac{\pi\Lambda}{4} \left( - g(\tau,\tau )+ \int_{0}^{\tau}d\tau' g(\tau',\tau')    \right)
    \label{final-decoherence-function}
\end{eqnarray}
where $g(\tau,\tau) $ is defined in \eqref{gtautau} for $\tau' = \tau'' $. Observe $g(\tau,\tau) \geq 0$, since it is given by the trace of a squared symmetric operator. 
%Note also that $ D(\tau)$ vanishes as $\Lambda \rightarrow \infty$. 
We refer to Appendix \ref{sec: appendix} for details. 

Since $ g(\tau,\tau) \geq 0 $, we have
\begin{eqnarray}
    D(\tau) \leq D_{0} \frac{\pi\Lambda}{4} \int_{0}^{\tau} d\tau' g(\tau',\tau')
    \label{inequality1}
\end{eqnarray}
If we can place a bound on $ g(\tau,\tau) \leq \gamma $, where $ \gamma $ is some positive constant, we can then bound the decoherence function as 
\begin{eqnarray}
    D(\tau) \leq D_{0} \frac{\pi\Lambda}{4}  \gamma \tau
    \label{inequality2}
\end{eqnarray}
Let us now place such bound on $ g(\tau,\tau) $.

Using \eqref{gtautau}, \eqref{delaQ} and \eqref{BH_quadrupole} we may write $g(\tau,\tau)$ explicitly in terms of $\hat{J}$ and its derivatives,
\begin{eqnarray}
    g(\tau,\tau) &=&  \ddot{\hat{Q}}^{B\nu}_{ij} \ddot{\hat{Q}}^{B\nu'  ij} \nonumber \\
    &=&  2 \left(  \ddot{\hat{J}}^{\nu} \cdot \ddot{\hat{J}}^{\nu'}  \right)  \left(  \hat{J}^{\nu} \cdot \hat{J}^{\nu'} \right) \nonumber \\
    &+&  4 \left(  \ddot{\hat{J}}^{\nu} \cdot \dot{\hat{J}}^{\nu'}  \right) \left(  \hat{J}^{\nu} \cdot \dot{\hat{J}}^{\nu'} \right) \nonumber \\
    &+& 2 \left(  \ddot{\hat{J}}^{\nu} \cdot \hat{J}^{\nu'}  \right) \left(  \hat{J}^{\nu} \cdot \ddot{\hat{J}}^{\nu'} \right) \nonumber \\
    &+& 4 \left(  \dot{\hat{J}}^{\nu} \cdot \ddot{\hat{J}}^{\nu'}  \right) \left(  \dot{\hat{J}}^{\nu} \cdot \hat{J}^{\nu'} \right) \nonumber \\
    &+& 4 \left(  \dot{\hat{J}}^{\nu} \cdot \dot{\hat{J}}^{\nu'}  \right)^{2} 
  \label{explicit-gtautau}
\end{eqnarray}
where sum over the superposition indices $\nu,\nu' = \pm$ is implied. Now, observe that the norm of $\hat{J}^{\nu}$ and its derivatives are bounded. Since $\vert \hat{J}^{\nu}\vert = 1$, we obtain from the equation of motion \eqref{eq-motion-final} that
\begin{eqnarray}
     \vert \dot{\hat{J}}^{\nu} \vert = \vert (\hat{\mathcal{E}}^{A\nu}\hat{J}^{\nu}) \vert  \leq \lambda_{1}^{\nu}
 \end{eqnarray}
 where $\lambda_{1}^{\nu} $ is the largest eigenvalue of $\hat{\mathcal{E}}^{A\nu}$. Similarly, we may also use \eqref{eq-motion-final} to write 
 \begin{eqnarray}
     \ddot{\hat{J}}^{\nu} = (\hat{\mathcal{E}}^{A\nu}\hat{J}^{\nu}) \times \left(  (\hat{\mathcal{E}}^{A\nu}\hat{J}^{\nu})\times \hat{J}^{\nu} \right) \nonumber \\ + \left(\hat{\mathcal{E}}^{A\nu}\left(   (\hat{\mathcal{E}}^{A\nu}\hat{J}^{\nu})\times \hat{J}^{\nu}\right)\right)\times \hat{J}^{\nu} 
 \end{eqnarray}
which implies,
  \begin{eqnarray}
     \vert \ddot{\hat{J}}^{\nu} \vert \leq 2(\lambda_{1}^{\nu})^{2}
 \end{eqnarray}
Combining these bounds with \eqref{explicit-gtautau} we find
\begin{eqnarray}
    g(\tau,\tau) \leq \gamma \equiv  36\left[(\lambda_{1}^{+})^{2} - (\lambda_{1}^{-})^{2} \right]^{2} \label{eq: gamma bound} \ .
\end{eqnarray}
Since the bound is a function of $\hat{\mathcal{E}}^{A\nu}(\psi,\theta,\phi)$ we have $\gamma = \gamma(\psi,\theta,\phi)$. Figure \ref{fig-final} shows a plot of $\gamma(\psi,\theta,\phi)$ (abbreviated as $\gamma(\theta,\phi)$) for fixed  $\psi = \pi/4$ as a function of the position of Bob's BH. We have $\mathrm{max} \ \gamma \approx 2\times10^{8} $. Despite reasonably large, we will see this represents a minuscule contribution when compared to other factors present in the decoherence function.

Finally, we can rewrite \eqref{inequality2} for $\tau = \omega T_{B}$ as,
\begin{eqnarray}
    D(T_{B}) \leq \frac{\gamma}{32} \chi^{8} \left(\frac{\ell_{P}}{b}\right)^{6} \left( \frac{R_{s}}{b}\right)^{9}  \left(  \frac{\bar{Q}_{A}}{b} \right)^{4}\left( \frac{T_{B}} {b}\right)
\end{eqnarray}
Using the condition that Alice does not decohere, this becomes,
\begin{eqnarray}
    D(T_{B}) \lesssim \frac{\chi^{8}}{32}\ \left( \frac{\sqrt{\gamma}\ell_{P}}{b}\right)^{2}  \left( \frac{R_{s}}{b}\right)^{9}  \left(  \frac{T_{A}}{b} \right)^{8} \left( \frac{T_{B}} {b}\right) 
\end{eqnarray}
Now, both the effective quantum field theory description of Einstein's gravity and the Newtonian limit require $\ell_{P}, R_{s} \ll b$, which implies the decoherence function is bounded by a number which is much smaller than one \footnote{Strictly speaking we have that $\sqrt{\gamma}\ell_{P} \approx 10^{-31}$ m, which is also much smaller than the value of $ b $ as required by effective field theory and the Newtonian limit.}, unless the conditions for spacelike separation \eqref{Alice_condition_spacelike} and \eqref{Bob_condition_spacelike} are violated. Quantum fluctuations in the gravitational waves prevent Alice and Bob from violating quantum mechanics or relativity!

%Incidentally, we have found that the gravitational decoherence due to distant Newtonian sources of gravity is totally negligible, due to the presence of small factors such as $(\sqrt{\gamma}\ell_{P}/b)^{2}$ in the decoherence function. 
%To have an idea on how small decoherence is, we may consider a binary BH system in a superposition of distinct orbits... estimate the decoherence rate due to the presence of a distant mass...
%Macroscopic quantum experiments seeking to observe massive superpositions are safe from distant black holes and other compact objects, even when the masses participating in the superposition are astronomical.

\section{Discussion}\label{Discussion}

We have considered a modified version of the gedankenexperiment discussed by Belenchia et al., where Alice and Bob attempt to communicate faster than light using massive quantum superpositions and gravitational waves emitted by the motion of a rotating black hole as a detector of gravitational superpositions. Similar to Belenchia et al., we conclude that quantum noise in gravitational waves as predicted by the effective quantum field theory description of Einstein's gravity is sufficient to prevent signaling and violations of quantum complementarity. This suggests that consistency of relativity with quantum mechanics requires the quantization of gravitational waves in the same manner as electromagnetic fields are quantized, even if the waves have been 
generated in situations where the gravitational field is strong, such as the motion of a rotating black hole. We can expect that some of the perturbations excited by the torqued precession considered in this work are described in terms of BH quasinormal modes. This points towards the need for quantizing black hole oscillations, which can exhibit significant self-interactions \cite{mitman2023nonlinearities, cheung2023nonlinear}. 

%The torqued precession which causes the time-dependent quadrupole moment, acts as a drive which excites metric perturbations near the BH. Heuristically, some of which we can expect to describe in terms of QNMs. This points towards the need for quantizing QNMs. The nonlinear nature of QNMs, plus their quantum nature, could lead to interesting quantum effects in GW signals.

Unambiguous direct detection of the quantum nature of gravitational waves is a near impossible task if one uses weakly gravitating systems, such as electromagnetic \cite{carney2024graviton}, interferometric \cite{guerreiro2025nonlinear}, or bar detectors \cite{manikandan2025testing}. However, strongly gravitating systems such as rotating black holes and other compact sources of highly curved spacetimes are likely to be sensitive probes of the quantum nature of gravitational waves \cite{guerreiro2025nonlinear}. 
%The quantum nature of strong-field gravitational waves, e.g. near the horizon of rotating black holes and other compact sources of highly curved spacetimes, will surely have a number of potentially observable consequences. 
Strong nonlinearities as those implied by Einstein's theory in highly curved backgrounds typically lead to non-classical radiative states, such as squeezing and entanglement \cite{Guerreirob2025} (see also \cite{das2025squeezed}), which may lead to detectable quantum signatures in gravitational waveforms originating from astrophysical sources. Besides being of clear fundamental importance, the inclusion of quantum effects of gravitational waves and fields might turn out to be essential for gravitational wave astronomy.

%In conclusion, quantum field fluctuations in GW coherent states are sufficient to prevent violations of causality and complementarity in our modified version of the Belenchia et al. gedankenexperiment. The waves emitted by the motion of Bob's BH are sourced by infinitesimal motion of the hole's angular momentum, and therefore originate from near the BH's surface. Consistency of the Newtonian limit of general relativity with the weak-field quantization of gravity requires the quantization of GWs originating from strongly curved spacetime backgrounds. 

%non-orthogonality, non-distinguishability... quantumness

%The dynamics of gravitational perturbations in strongly curved spacetimes can be highly nonlinear. \textcolor{red}{WORK IN PROGRESS}...

%This suggests that GWs must be quantized and undergo quantum fluctuations even when they originate from perturbations of strongly curved spacetime backgrounds. 

%Gravitational waves bring out the news that Bob's BH is undergoing precession due to Alice's tidal field, i.e. these GWs are sourced by infinitesimal perturbations of the hole's angular momentum, according to Eq. \eqref{hartle-thorne-dimensionless}. 

%\begin{itemize}
    %\item If rotations of Kerr BHs turn out to be described in terms of quasinormal modes, can we extend the argument for quantization to all quasinormal modes?

    %\item Make the statement that quantum fluctuations + nonlinearities imply nonclassical effects (Mollow \& Glauber)
%\end{itemize}

\acknowledgments{
The authors acknowledge Bruno Suassuna, Carlos Tomei and Nami Svaiter for conversations. T.G. acknowledges Roberto Cisne, whose essential support made this work possible. We acknowledge support from the Coordenac\~ao de Aperfei\c{c}oamento de Pessoal de N\'ivel Superior - Brasil (CAPES) - Finance Code 001, Conselho Nacional de Desenvolvimento Cient\'ifico e Tecnol\'ogico (CNPq), Funda\c{c}\~ao de Amparo \`a Pesquisa do Estado do Rio de Janeiro (FAPERJ Scholarship No. E-26/200.251/2023, E-26/210.249/2024 and FAPERJ PhD merit fellowship - FAPERJ Nota 10, 203.709/2025), Funda\c{c}\~ao de Amparo \`a Pesquisa do Estado de São Paulo (FAPESP processo 2021/06736-5), the Serrapilheira Institute (grant No. Serra – 2211-42299) and StoneLab.}

\bibliography{main}

\onecolumngrid
\appendix

%=============================================================
\newpage
\section{Tidal field and decoherence function}\label{sec: appendix}

\subsection{Tidal field}

Using the definition of $\hat{\mathcal{E}}^{A}$ provided in the main text (Eq. \eqref{dimensionless-tidal-field}), we may evaluate the tidal field felt by Bob's BH at an arbitrary position $\hat{\mathbf{r}} = (x,y,z)$, where $ x = \sin\theta \cos\phi,  y = \sin\theta\sin\phi, z = \cos\theta  $. Define,
\begin{eqnarray}
    \mathcal{Q} = \hat{Q}^{A}_{xx} x^{2} + \hat{Q}^{A}_{yy} y^{2} + 2 \hat{Q}^{A}_{xy}xy + \hat{Q}^{A}_{zz} z^{2}.
\end{eqnarray}
Direct application of \eqref{dimensionless-tidal-field} then gives,
\begin{eqnarray}
    \hat{\mathcal{E}}^{A}_{xx} &=& \frac{3}{2}\left[  \hat{Q}^{A}_{xx}  - 10 (\hat{Q}^{A}_{xx} x + \hat{Q}^{A}_{xy}y)x + \frac{35}{2}\mathcal{Q} x^{2}\right], \\ 
    \hat{\mathcal{E}}^{A}_{xy} &=& \frac{3}{2}\left[  \hat{Q}^{A}_{xy}  - 5 \left( (\hat{Q}^{A}_{yy} y + \hat{Q}^{A}_{xy}x)x 
 + (\hat{Q}^{A}_{xx} x + \hat{Q}^{A}_{xy}y)y\right) + \frac{35}{2}\mathcal{Q} xy\right], \\
    \hat{\mathcal{E}}^{A}_{xz} &=& \frac{3}{2}\left[  -5 ((\hat{Q}^{A}_{xx}x + \hat{Q}^{A}_{xy}y)z + (\hat{Q}^{A}_{zz}z)x) + \frac{35}{2}\mathcal{Q} xz\right], \\
    \hat{\mathcal{E}}^{A}_{yy} &=& \frac{3}{2}\left[  \hat{Q}^{A}_{yy}  - 10 (\hat{Q}^{A}_{yy} y + \hat{Q}^{A}_{xy}x)y + \frac{35}{2}\mathcal{Q} y^{2}\right], \\
    \hat{\mathcal{E}}^{A}_{yz} &=& \frac{3}{2}\left[  -5 \left(  (\hat{Q}^{A}_{yy} y + \hat{Q}^{A}_{xy}x)z + (\hat{Q}^{A}_{zz}z) y  \right) + \frac{35}{2}\mathcal{Q} yz\right], \\
    \hat{\mathcal{E}}^{A}_{zz} &=& \frac{3}{2}\left[  \hat{Q}^{A}_{zz}  - 10 \hat{Q}^{A}_{zz}z^{2} + \frac{35}{2}\mathcal{Q} z^{2}\right].
\end{eqnarray}

\subsection{Eigenvalues}

%Since each element of Alice's quadrupole is given by Eqs. \eqref{eq: Qxx}-\eqref{eq: Qzz}, 

Direct calculation shows that the eigenvalues of $\mathcal{E}^A(\psi,\theta,\phi)$ are given by
\begin{comment}
\begin{eqnarray}
    && \lambda_i^3 + \frac{45}{8}\lambda_i^2 \left(1 + 3\cos2\theta - 6 \cos2(\phi-\psi) \sin^2\theta \right) \\ 
    &&- \frac{27}{128}\lambda_i\left( 1003 - 420 \cos2\theta - 135 \cos4\theta + 240(5 + 3 \cos2\theta) \sin^2\theta \cos2(\phi-\psi)  - 360\sin^4\theta \cos4(\phi-\psi)\right) \nonumber \\
    &&-\frac{27}{128} \left(3121 - 540\cos2\theta -405\cos4\theta+540(1-\cos4\theta) \cos2(\phi-\psi)-1080\sin^4\theta \cos4(\phi-\psi)  \right) = 0 \nonumber
\end{eqnarray}
where $\lambda_2$ is the trivial solution given by $\lambda_2 = -3$. The other two satisfy the following inequality:
\begin{equation}
    \lambda_1 >0 > \lambda_2 \geq \lambda_3,
\end{equation}
and are given by
\end{comment}
\begin{equation}\label{eq: eigenvalues}
    \lambda_1 = \frac{3}{64}\left( -A +3\sqrt{2B} \right), \lambda_{2} = -3, ~\lambda_3 = -\frac{3}{64}\left( A +3\sqrt{2B} \right),
\end{equation}
where
\begin{eqnarray}
    A&=& 28 + 180\left[ \cos2\theta - \cos2(\phi-\psi)\right] + 90\left[ \cos2(\theta + \phi-\psi) + \cos2(\theta - \phi + \psi)\right], \label{eq: A factor}\\
    B&=& 7842 - 2200\left[ \cos2\theta - \cos2(\phi-\psi)\right] +630\left[ \cos4\theta + \cos4(\phi-\psi)\right] \label{eq: B factor} \nonumber \\
    &&  - 3040\cos2\theta \cos2(\phi-\psi) + 210 \cos4\theta \cos4(\phi-\psi) - 840\left[ \cos2\theta \cos4(\phi - \psi) - \cos4\theta \cos2(\phi-\psi)\right]
\end{eqnarray}
and 
\begin{equation}
    \lambda_1 >0 > \lambda_2 \geq \lambda_3.
\end{equation}

Note that for $\theta = n\pi/2$ and $\phi = m\pi/2$ where $n,m \in \mathbb{Z}$, $A$ and $B$ can be written as
\begin{eqnarray}
    A&=& 28 + 180\left[ (-1)^{n} - (-1)^{m}\cos2\psi\right] + 90\left[ (-1)^{n+m} + (-1)^{n-m}\right]\cos2\psi, \\
    B&=& 7842 - 2200\left[ (-1)^{n} - (-1)^{m}\cos2\psi\right] +630\left[ 1 + \cos4\psi\right] \nonumber \\
    &&  - 3040(-1)^{n+m} \cos2\psi + 210  \cos4\psi - 840\left[ (-1)^{n} \cos4 \psi -  (-1)^{m}\cos2\psi\right],
\end{eqnarray}
and we see that $\lambda_1$ and $\lambda_3$ become even functions of $\psi$.

%Since we define $\lambda_1^{\pm} = \lambda_1(\theta,\phi,\pm \psi)$, we can obtain $\gamma$, defined by Eq. \eqref{eq: gamma bound}, using Eqs. \eqref{eq: eigenvalues}-\eqref{eq: B factor}. 

\subsection{Decoherence function}

Eq. \eqref{coherent_state_integration} in the main text can be cast in the dimensionless form  $ D(t) = D_{0} F(\Lambda,\tau) $
where $D_{0} $ is defined in \eqref{D0} and
\begin{equation}
    F(\Lambda,\tau) = \int_{0}^{\Lambda}dz ~z \int_{0}^{\tau}\int_{0}^{\tau} d\tau' d\tau'' \sin z(\tau-\tau') \sin z(\tau-\tau'') g(\tau',\tau''),
    \label{dimensionless-function}
\end{equation}
with $ z = k/\omega$ and $g(\tau',\tau'')$ as defined in \eqref{gtautau} with rescaled time $t \rightarrow \omega t = \tau$. Let us perform the integral in $k$ in \eqref{dimensionless-function}. We have,
\begin{eqnarray}
    \int_{0}^{\Lambda} dz  \ z  \sin(az) \ \sin(bz) = \frac{1}{2}   \  \underbrace{\left(\frac{\Lambda\sin(\Lambda(a-b))}{(a-b)} + \frac{\cos(\Lambda(a-b)) - 1}{(a-b)^{2}}\right)}_{\text{(I)}} 
    - \frac{1}{2} \underbrace{\left(\frac{\Lambda\sin(\Lambda(a+b))}{(a+b)} + \frac{\cos(\Lambda(a+b)) - 1}{(a+b)^{2}} \right)}_{\text{(II)}}, \nonumber \\
    \label{k-integral1}
\end{eqnarray}
where $a = \tau - \tau'$ and $ b = \tau - \tau''$. Note that $a-b = \tau'' - \tau'$ and $a+b = 2\tau - \tau'-\tau''$. Let us proceed term by term in \eqref{k-integral1}.

\textbf{Term (I)} gives, in the limit of large cut-off,
\begin{eqnarray}
    \frac{1}{2}\left(\frac{\Lambda\sin(\Lambda(a-b))}{(a-b)}\right) &\rightarrow& \frac{\pi \Lambda}{2} \delta(\tau'' - \tau'), \\
    \frac{1}{2}\left(\frac{\cos(\Lambda(a-b)) - 1}{(a-b)^{2}}\right) &\rightarrow& -\frac{\pi \Lambda}{4} \delta(\tau'' - \tau').
\end{eqnarray}
This contributes to $F(\Lambda,\tau) $ as 
\begin{eqnarray}
    \text{(I)} = \frac{\pi\Lambda}{4}\int_{0}^{\tau}d\tau' g(\tau',\tau').
    \label{term-I}
\end{eqnarray}

Similarly, \textbf{Term (II)} gives
\begin{eqnarray}
    \frac{1}{2}\left(\frac{\Lambda\sin(\Lambda(a+b))}{(a+b)} \right) &\rightarrow&  \frac{\pi\Lambda}{2} \delta(\tau'' - (2\tau-\tau')), \\
    \frac{1}{2}\left(\frac{\cos(\Lambda(a+b)) -1}{(a+b)({2}} \right) &\rightarrow&  -\frac{\pi\Lambda}{4} \delta(\tau'' - (2\tau-\tau')),
    \label{delta2}
\end{eqnarray}
and we have
\begin{eqnarray}
    \text{(II)} &=& \frac{\pi\Lambda}{4}\int_{0}^{\tau}\int_{0}^{\tau} d\tau' d\tau'' \delta(\tau'' - (2\tau-\tau')) g(\tau',\tau'') \nonumber \\
    &=& \frac{\pi\Lambda}{4} \int_{0}^{\tau} d\tau' \int_{-\infty}^{\infty} d\tau'' H(\tau'')H(\tau-\tau'') \delta(\tau'' - (2\tau-\tau')) g(\tau',\tau'') \nonumber \\
    &=& \frac{\pi\Lambda}{4}\int_{0}^{\tau} d\tau'  H(2\tau-\tau')H(\tau'-\tau) g(\tau',2\tau - \tau') \nonumber \\ 
    &=& \frac{\pi\Lambda}{4}\int_{0}^{\tau} d\tau'  H(\tau'-\tau) g(\tau',2\tau - \tau') \nonumber \\
    &=& \frac{\pi\Lambda}{4}H(0) g(\tau,\tau) \sim \frac{\pi\Lambda}{4} g(\tau,\tau). 
\end{eqnarray}
Therefore, in the limit of large $\Lambda$,
\begin{eqnarray}
    F(\Lambda,\tau) \approx \frac{\pi\Lambda}{4} \left( - g(\tau,\tau )+ \int_{0}^{\tau}d\tau' g(\tau',\tau')    \right).
\end{eqnarray}

\end{document}